\documentclass[%
 reprint,
 amsmath,amssymb,
 aps,
prl]{revtex4-1}

\usepackage{subfigure}
\usepackage{graphicx}
\usepackage{dcolumn}% Align table columns on decimal point
\usepackage{bm}% bold math
\begin{document}

\preprint{APS/123-QED}

\title{Magnon Spin-Momentum Locking: \\Various Spin Vortices and Dirac Magnons in Noncollinear Antiferromagnets}% Force line breaks with \\

\author{Nobuyuki Okuma}
\email{okuma@hosi.phys.s.u-tokyo.ac.jp}
 %\altaffiliation{Department of Physics, University of Tokyo, Hongo 7-3-1, 113-0033, Japan}%Lines break automatically or can be forced with \\
\affiliation{%
 Department of Physics, University of Tokyo, Hongo 7-3-1, Tokyo 113-0033, Japan
}%

\if0%
\author{Charlie Author}
 \homepage{http://www.Second.institution.edu/~Charlie.Author}
\affiliation{
 Second institution and/or address\\
 This line break forced% with \\
}%
\affiliation{
 Third institution, the second for Charlie Author
}%
\author{Delta Author}
\affiliation{%
 Authors' institution and/or address\\
 This line break forced with \textbackslash\textbackslash
}%

\collaboration{CLEO Collaboration}%\noaffiliation
\fi%
\date{\today}% It is always \today, today,

\begin{abstract} 
We generalize the concept of the spin-momentum locking to magnonic systems
and derive the formula to calculate the spin expectation value for one-magnon states of general two-body spin Hamiltonians.
We give no-go conditions for magnon spin to be independent of momentum.
As examples of the magnon spin-momentum locking, we analyze a one-dimensional antiferromagnet with the N\'eel order and two-dimensional kagome lattice antiferromagnets with the 120$^\circ$ structure. 
We find that the magnon spin depends on its momentum even when the Hamiltonian has the $z$-axis spin rotational symmetry, 
which can be explained in the context of a singular band point or a $U(1)$ symmetry breaking.
A spin vortex in momentum space generated in a kagome lattice antiferromagnet has the winding number $Q=-2$, while the typical one observed in topological insulator surface states is characterized by $Q=+1$.
A magnonic analogue of the surface states, the Dirac magnon with $Q=+1$, is found in another kagome lattice antiferromagnet.
We also derive the sum rule for $Q$ by using the Poincar\'e-Hopf index theorem.

\begin{description}
\item[PACS numbers]
75.30.Ds,75.50.Ee,72.20.-i, 75.76.+j,85.70.-w
\end{description}

\end{abstract}

\maketitle

%\tableofcontents
{\it Introduction}.\textemdash The physics of magnons \cite{chumak,magnonics}, the quanta of spin wave excitations, is enriched by their multiband nature.
Although magnons have no internal degrees of freedom other than spin, such as atomic orbitals, the presence of chemical and magnetic sublattices allows magnonic systems to exhibit nontrivial band structures.
In particular, recent studies have generalized many concepts in topological band theory \cite{hasan,xlq,niu},
established in multiband electron systems,
to magnonic systems, e.g. 
the magnon Hall effect \cite{fujimoto,katsura,onose,owerre3}, magnon topological insulators \cite{shindou,lzhang,chisnell,owerre}, and Weyl \cite{fyli} (Dirac \cite{fransson}) magnons.

%More recently, magnon spin angular momentum in multi-band systems has also been studied in addition to their topological properties \cite{rcheng2, zyuzin,ohnuma,rcheng}. 
Another interesting feature is spin angular momentum carried by one magnon.
Magnon spin in multiband systems, which has been studied for a long time \cite{kittel}, has attracted renewed attention \cite{rcheng2, zyuzin,ohnuma,rcheng} motivated by recent developments in spintronic techniques \cite{chumak}.
Except for in simple collinear ferromagnets, magnon spin generally depends on its band properties.
For instance, the magnon spin Nernst effect in antiferromagnets, which is a magnonic analogue of the spin Hall effect, is interpreted as the two copies of the magnon Hall effect for magnons with opposite spins \cite{rcheng2, zyuzin}.
This example shows the importance of considering the spin direction of each magnon mode in magnon spintronics \cite{chumak}.

\begin{figure}[]
\begin{center}
\includegraphics[bb=0 0 4076 2559,width=7cm,angle=0,clip]{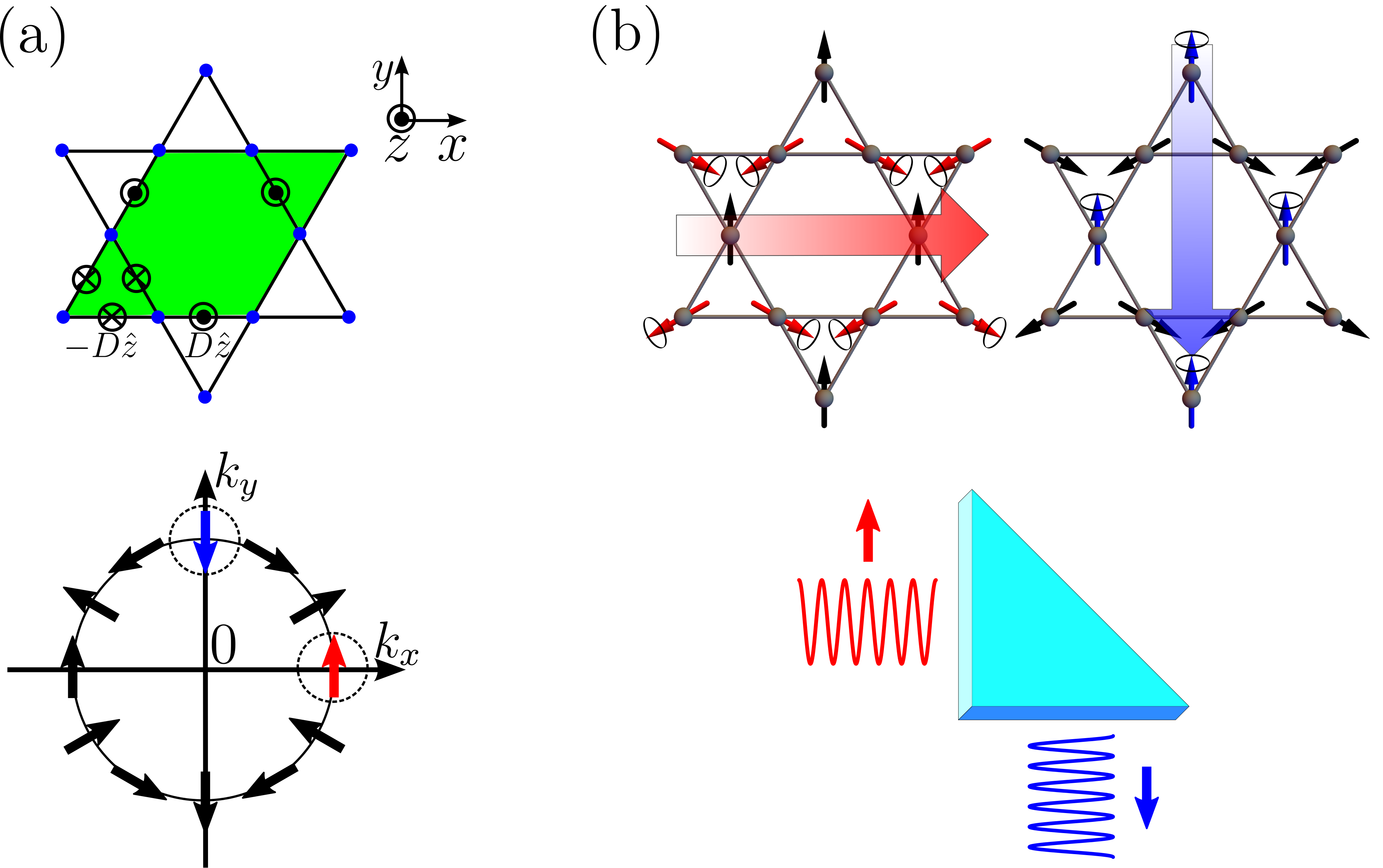}
\caption{(a) Schematics of the magnon spin-momentum locking with the winding number $Q=-2$ in a magnon band of a kagome lattice antiferromagnet with the Dzyaloshinskii-Moriya interaction denoted by $\pm D\hat{z}$. The magnetic unit cell, shown by the green region, is the same as the chemical unit cell. Precise vector plots of magnon spin are shown in Fig. \ref{fig3}. 
(b) Upper: real space illustration of spin-momentum-locked magnon modes with $\bm{k}=(k_x,0)$ (red) and $\bm{k}=(0,-k_y)$ (blue).
 Lower: schematics of magnon spin flip device using the spin-momentum locking with $Q=-2$.}
\label{fig1}
\end{center}
\end{figure}

In this Letter, we generalize the concept of spin-momentum locking to magnonic systems.
The conventional spin-momentum locking, in which electron spin depends on its momentum, is described by a noninteracting Hamiltonian without rotational symmetry in spin space, 
such as the Dirac Hamiltonian of topological insulator surface states \cite{hasan,xlq}.
We define the magnon spin for each band and prove a no-go theorem which states that magnon spin is momentum independent for several conditions.
By performing a numerical calculation for kagome lattice antiferromagnets with a 120$^\circ$ structure, 
we find a spin-momentum-locked magnon band characterized by the winding number \cite{winding} $Q=-2$ [Fig. $\ref{fig1}$].
Our results for spin Hamiltonians with $z$-axis spin rotational symmetry demonstrate for the first time that spin-momentum locking can be generated through spontaneous symmetry breaking.

{\it Definition of $\bm{k}$-dependent magnon spin}.\textemdash 
A general two-body spin interaction Hamiltonian is given by
\begin{align}
H=\frac{1}{2}\sum_{\bm{R},\bm{R'}}\sum_{i,j=1}^N\sum_{a,b}J^{ab}_{ij}(\bm{R},\bm{R'})S^{a}_{\bm{R},i}S^b_{\bm{R'},j},\label{spinham}
\end{align}
where $\bm{S}_{\bm{R},i}=(S^x_{\bm{R},i},S^y_{\bm{R},i},S^z_{\bm{R},i})$ is the spin operator at each site, $\bm{R},\bm{R'}$ denote the magnetic lattice vectors, $i,j$ denote the magnetic-sublattice indices, and $N$ is the number of sites in a magnetic unit cell.
To rewrite the spin Hamiltonian ($\ref{spinham}$) in terms of spin excitations (magnons) around a classical ground state, we introduce the Holstein-Primakoff boson operators ($a,a^{\dagger}$),
\begin{align}
\bm{S}_{\bm{R},i}\simeq&\bm{M}^z_i(S_0-a^{\dagger}_{\bm{R},i}a_{\bm{R},i})+\bm{M}^x_i\sqrt{2S_0}\frac{a_{\bm{R},i}+a^{\dagger}_{\bm{R},i}}{2}\notag\\
&+\bm{M}^y_i\sqrt{2S_0}\frac{a_{\bm{R},i}-a^{\dagger}_{\bm{R},i}}{2i},\label{holsteinspin}
\end{align}
where $S_0$ is the size of the spin and $\{\bm{M}_i^a\}$ is the set of the basis vectors of the rotating frame in which $\bm{M}_i^{z}$ is in the direction of classical spin at $i$.
The corresponding quadratic form of boson Hamiltonian is given by
\begin{align}
H=\frac{1}{2}\sum_{\bm{k}}(\bm{a}_{\bm{k}}^{\dagger},\bm{a}_{-\bm{k}})\cdot \hat{\mathcal{H}}_{\bm{k}}\cdot\binom{\bm{a}_{\bm{k}}}{\bm{a}_{-\bm{k}}^{\dagger}},\label{bosonham}
\end{align}
where $\bm{k}$ is the crystal momentum, $\bm{a}^{\dagger}_{\bm{k}}=(a^{\dagger}_{\bm{k},1},\cdots,a^{\dagger}_{\bm{k},N})$, and $\hat{\mathcal{H}}_{\bm{k}}$ is a $2N\times2N$
bosonic Bogoliubov-de Gennes Hamiltonian.
We ignore magnon-magnon interactions in Eqs. ($\ref{holsteinspin}$) and ($\ref{bosonham}$) by assuming $S_0\gg 1$ \cite{magmageffect}.
The eigenenergy problem of Eq. (\ref{bosonham}) can be solved by the Bogoliubov transformation \cite{colpa},
\begin{align}
&\hat{Q}^{\dagger}_{\bm{k}}\hat{\mathcal{H}}_{\bm{k}}\hat{Q}_{\bm{k}}=
\begin{pmatrix}
\hat{E}_{\bm{k}} & 0\\
0 & \hat{E}_{-\bm{k}}
\end{pmatrix},\notag\\
&H=\sum_{\bm{k},\alpha}E_{\bm{k},\alpha}b^{\dagger}_{\bm{k},\alpha}b_{\bm{k},\alpha},
\end{align}
where $\hat{Q}_{\bm{k}},\hat{Q}^{\dagger}_{\bm{k}}$ are $2N\times2N$ paraunitary matrices and $\hat{E}_{\bm{k}}=\mathrm{diag}(E_{\bm{k},1},\cdots,E_{\bm{k},\alpha},\cdots, E_{\bm{k},N})$.
($b,b^{\dagger}$) are the magnon field operators, which satisfy
\begin{align}
a_{\bm{k},i}=[\hat{Q}_{\bm{k}}]_{i,\alpha}b_{\bm{k},\alpha}+[\hat{Q}_{\bm{k}}]_{i,\alpha+N}b^{\dagger}_{-\bm{k},\alpha}.\label{bogo}
\end{align}
Using Eqs. (\ref{holsteinspin}) and (\ref{bogo}), the total spin operator is given by
\begin{align}
\bm{S}_{\mathrm{tot}}=&\sum_{\bm{R}}\sum_{i}\bm{S}_{\bm{R},i}\notag\\
=&\sum_{\bm{k},\alpha}\left[\sum_{i}(-\bm{M}^z_i)\left\{|[\hat{Q}_{\bm{k}}]_{i,\alpha} |^2+|[\hat{Q}_{-\bm{k}}]_{i,\alpha+N} |^2\right\}\right] b^{\dagger}_{\bm{k},\alpha}b_{\bm{k},\alpha}\notag\\
&+(\mathrm{off}\mathchar`-\mathrm{diagonal\ terms})\notag\\
&+(\mathrm{zeroth\mathchar`-\ and\ first\mathchar`-order\ terms\ of\ }b,b^{\dagger}).
\end{align}
Thus, the $\bm{k}$-dependent magnon spin carried by a one-magnon state $|\bm{k},\alpha\rangle\equiv b^{\dagger}_{\bm{k},\alpha}|0\rangle$, where $|0\rangle$ is the Fock vacuum of ($b,b^{\dagger}$), is given by
\begin{widetext}
\begin{align}
\bm{S}_{\bm{k},\alpha}\equiv\langle\bm{k},\alpha|\bm{S}_{\mathrm{tot}}  |\bm{k},\alpha\rangle-\langle0|\bm{S}_{\mathrm{tot}}  |0\rangle
=\sum_{i}(-\bm{M}^z_i)\left\{|[\hat{Q}_{\bm{k}}]_{i,\alpha} |^2+|[\hat{Q}_{-\bm{k}}]_{i,\alpha+N} |^2\right\}.\label{mainresult}
\end{align}
\end{widetext}

{\it No-go conditions for ordered magnets}.\textemdash 
%For several models, we can prove that there is no $\bm{k}$-dependence of the magnon spin defined in Eq. (\ref{mainresult}).
In noninteracting electron systems, spin-momentum locking is forbidden for Hamiltonians with rotational symmetries \cite{rotsym}.
In magnonic systems, however, it does not hold, because of spontaneous symmetry breaking.
In the following, we write no-go conditions for magnon spin to be independent of momentum.
We first consider two-body spin Hamiltonians with $SO(3)$-rotational symmetry in spin space such as isotropic Heisenberg models.
Suppose that $|0\rangle$ is a ground state with a spontaneous symmetry breaking. 
Using $[H,S_{\mathrm{tot}}^a]=0$, we can rewrite $S^a_{\mathrm{tot}}$ in terms of $(b,b^{\dagger})$ up to the second order \cite{supplemental},
\begin{align}
S^a_{\mathrm{tot}}=&(\mathrm{const)}+\sum_{\bm{k},\alpha}S^a_{\bm{k},\alpha}b^{\dagger}_{\bm{k},\alpha}b_{\bm{k},\alpha}\notag\\
&+\sum_{m=1}^{n_{\mathrm{NG}}} \left(C^a_mb_{\mathrm{NG},m} +(C^a_m)^*b^{\dagger}_{\mathrm{NG},m}   \right),\label{ngrepresentation}
\end{align}
where ($b_{\mathrm{NG},m},b^{\dagger}_{\mathrm{NG},m}$) are field operators of massless Nambu-Goldstone (NG) modes associated with the spontaneous symmetry breaking, $m$ denotes the index of the independent massless NG modes, $n_{\mathrm{NG}}$ is the total number of the massless NG modes, and $C^a_m$ are complex numbers.
For collinear magnets with the symmetry breaking: $SO(3)$$\rightarrow$$U(1)$, $|0\rangle$ is an eigenstate of the unbroken generator $\tilde{S}^z_{\mathrm{tot}}$, which does not include ($b_{\mathrm{NG},m},b^{\dagger}_{\mathrm{NG},m}$).
A one-magnon state $|\bm{k},\alpha\rangle$ is also an eigenstate of $\tilde{S}^z_{\mathrm{tot}}$:
\begin{align}
\tilde{S}^z_{\mathrm{tot}}b^{\dagger}_{\bm{k},\alpha}|0\rangle&=\left[(Const.)+\sum_{\bm{k},\beta}\tilde{S}^z_{\bm{k},\beta}b^{\dagger}_{\bm{k},\beta}b_{\bm{k},\beta}\right]b^{\dagger}_{\bm{k},\alpha}|0\rangle\notag\\
&\propto b^{\dagger}_{\bm{k},\alpha}|0\rangle.
\end{align}
Using the notation $|\bm{k},M\rangle$ that is an eigenstate of $\tilde{S}^z_{tot}$ with an eigenvalue M instead of $|\bm{k},\alpha\rangle$, we obtain 
\begin{align}
&\langle\bm{k},M|\tilde{S}^z_{\mathrm{tot}}|\bm{k},M\rangle=M,\notag\\
&\langle\bm{k},M|\tilde{S}^{x(,y)}_{\mathrm{tot}}  |\bm{k},M\rangle=\pm 1/i\langle\bm{k},M|[\tilde{S}^{y(,x)}_{\mathrm{tot}},\tilde{S}^z_{\mathrm{tot}}]  |\bm{k},M\rangle=0,\label{quantized}
\end{align}
where $\tilde{S}^{x(,y)}_{\mathrm{tot}} $ satisfy $[\tilde{S}^a_{\mathrm{tot}},\tilde{S}^b_{\mathrm{tot}}]=i\epsilon_{abc}\tilde{S}^c_{\mathrm{tot}}$.
Equation ($\ref{quantized}$) shows that $S^a_{\bm{k},\alpha}$ take quantized values.
Because quantized spin components are not changed under a small momentum change $\bm{k}\rightarrow\bm{k}+\delta\bm{k}$,
we cannot expect the $\bm{k}$-dependent magnon spin in isotropic Heisenberg models with collinear ground states.
Note that there is a $trivial$ exception.
The above statement assumes the smoothness of $S^a_{\bm{k},\alpha}$ on the magnon band $\alpha$.
However, when we cannot avoid a singularity such as a band crossing point in a one-dimensional system in the adiabatic deformation $\bm{k}\rightarrow \bm{k'}$,
$S^{a}_{\bm{k},\alpha}$ can be changed across the singular region.
For noncollinear and noncoplanar systems with symmetry breaking: $SO(3)$$\rightarrow \{e\}$, where $e$ is the identity element, we can also expect the $\bm{k}$-dependent magnon spin since there is no unbroken generator $\tilde{S}^z_{\mathrm{tot}}$ such that $\tilde{C}^z_m=0$ \cite{triangular}.

\begin{figure}[]
\begin{center}
\includegraphics[bb=0 0 787 661,width=7cm,angle=0,clip]{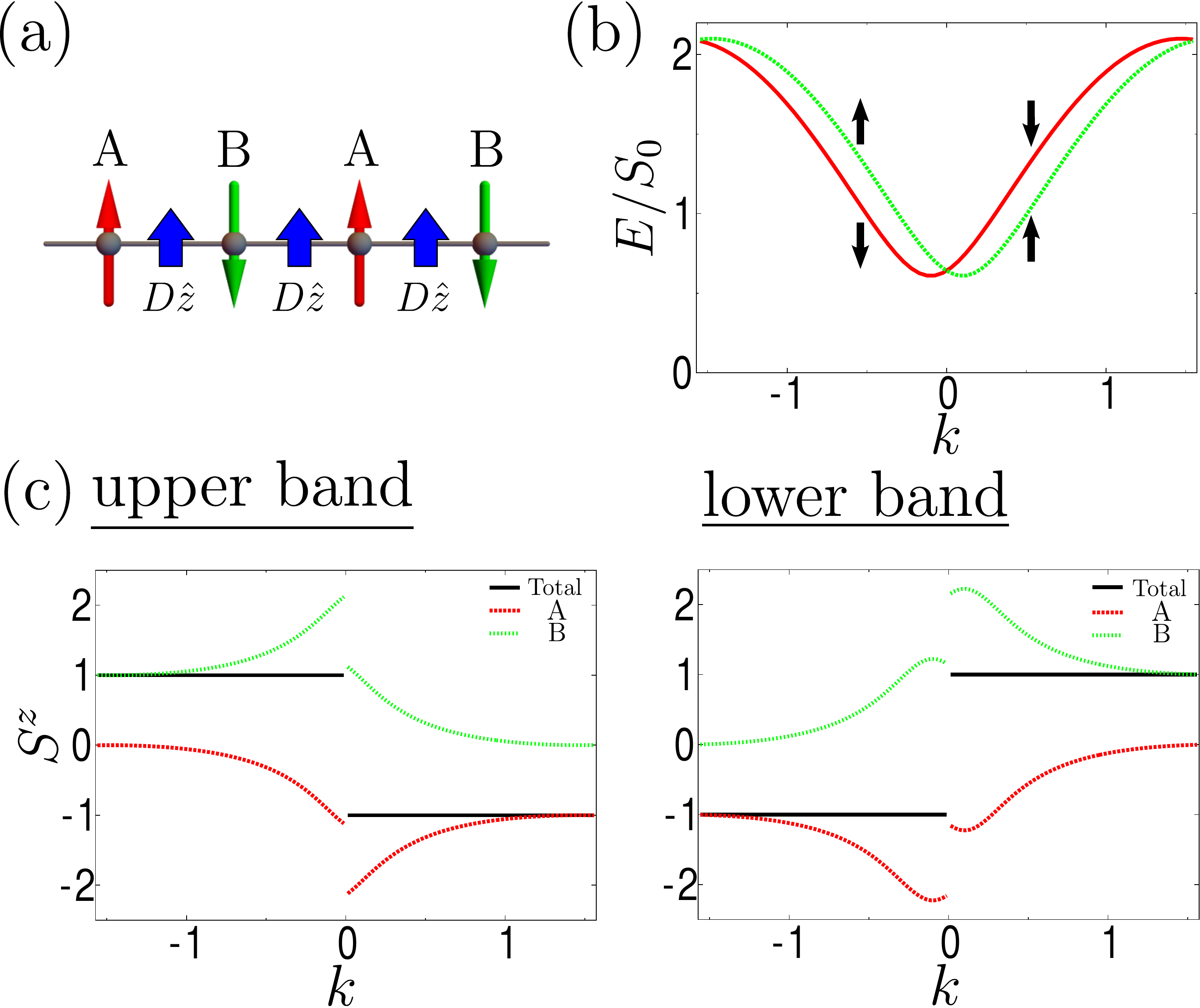}
\caption{(a) Schematics of the ground state of a one-dimensional antiferromagnet described by Eq. (\ref{1dantiferro}). 
The ground state is the N\'eel state with two sublattices ($A$, $B$), and their spins are parallel to the DM vector $D\hat{z}$.
(b) Magnon band dispersions for $J=1, D=0.1,$ and $K=-0.05$. The solid and dotted lines denote the magnon states with $S^z=-1$ and 1, respectively.
(c) The contributions from the $A$ and $B$ sublattices to the $z$-component spin in the upper and lower bands are plotted for the momentum $k$. The total $S^z$ is quantized, and its sign is changed across the band crossing points.}
\label{fig2}
\end{center}
\end{figure}

Next, we apply a similar argument to the Hamiltonians with the $U(1)$-rotational symmetry around the $z$ axis such as $XY$ models.
Since $[H,S^z_{\mathrm{tot}}]=0$ and $[H,S^{x,y}_{\mathrm{tot}}]\neq0$, Eq. ($\ref{ngrepresentation}$) holds only for the $z$-component spin.
When there is no massless mode, $C^z_m$ and $(C^{z}_m)^*$ in Eq. ($\ref{ngrepresentation}$) are zero. 
Thus, $|0\rangle$ and $|\bm{k},\alpha\rangle$ are eigenstates of $S^z_{\mathrm{tot}}$, and we cannot expect the $\bm{k}$-dependent magnon spin.
It is important to note that the above argument does not hold in the presence of the singularity discussed above and symmetry breaking: $U(1)$$\rightarrow$$\{e\}$ in which there is one massless NG mode \cite{watanabe}, or equivalently, the states are no longer the eigenstates of $S^{z}_{\mathrm{tot}}$.
We construct examples for both cases in the following parts.

\begin{figure*}[]
\begin{center}
\includegraphics[bb=0 0 5244 2006,width=14cm,angle=0,clip]{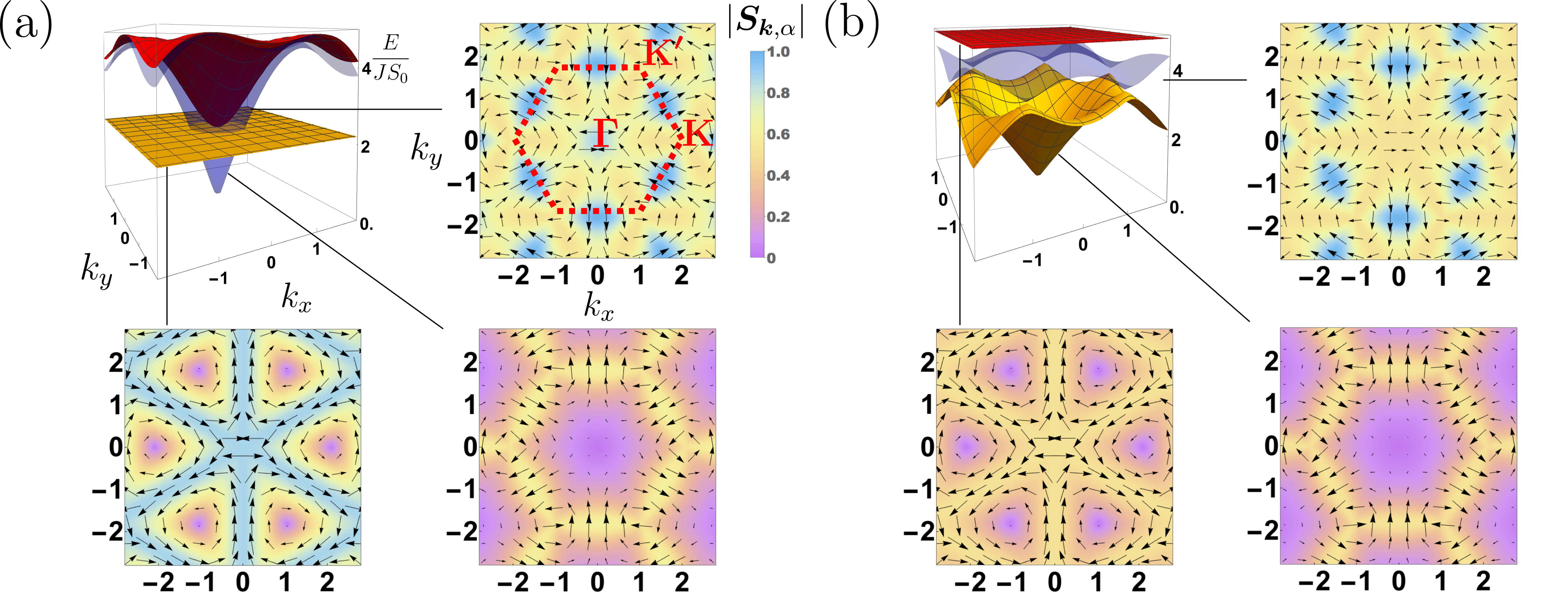}
\caption{Magnon band dispersions and $\bm{k}$-dependent spin in kagome lattice antiferromagnets described by Eq. ($\ref{kagomeham}$) 
with (a) $J^x=J^y=J^z=1,D=0.1$ and (b) $J^x=J^y=1,J^z=D=0$. The lattice constant $a=1$.  The hexagonal region surrounded by the red dotted lines is the first Brillouin zone.}
\label{fig3}
\end{center}
\end{figure*}

{\it Trivial example in 1D antiferromagnet}.\textemdash 
To gain some insight into the $\bm{k}$-dependent magnon spin,
we first consider a simple model of a one-dimensional antiferromagnet, which has been studied in the context of the spin wave field effect transistor \cite{rcheng},
\begin{align}
H^{1D}=\sum_{\langle i,j \rangle}\left[J\bm{S}_i\cdot\bm{S}_j+D\hat{z}\cdot\left(\bm{S}_i\times\bm{S}_j \right)\right]+K\sum_{i}S^2_{i,z},\label{1dantiferro}
\end{align}
where $J>0$ is the nearest-neighbor exchange coupling, $D$ is the strength of the Dzyaloshinskii-Moriya (DM) interaction, and $K<0$ is the easy-axis anisotropy.
Although the DM interaction and the anisotropy breaks the $SO(3)$ symmetry, they preserve the $U(1)$ symmetry around the $z$ axis.

For sufficiently small $D$, the classical ground state is the N\'eel state with two sublattices, $A$ with up spin and $B$ with down spin [Fig. $\ref{fig2}$(a)].
Using the Holstein-Primakoff transformation around the N\'eel state,
\begin{align}
&S^{\pm}_{R,A}=\sqrt{2S_0}a^{(\dagger)}_{R,A}, S^z_{R,A}=S_0-a^{\dagger}_{R,A}a_{R,A},\notag\\
&S^{\mp}_{R,B}=\sqrt{2S_0}a^{(\dagger)}_{R,B}, S^z_{R,B}=a^{\dagger}_{R,B}a_{R,B}-S_0,
\end{align} 
we can rewrite Eq. ($\ref{1dantiferro}$) in terms of magnons as
\begin{align}
H^{1D}=\frac{1}{2}\sum_{k}\Psi^{\dagger}_{k}
\begin{pmatrix}
X(k) & 0 & 0 & Y_{-}(k)\\
0 & X(k) & Y_{+}(k)& 0\\
0& Y_{+}(k)&X(k) & 0\\
Y_{-}(k)&0 &0 &X(k)
\end{pmatrix}\label{1dham}
\Psi_{k},
\end{align}
where $k$ is the one-dimensional momentum, $\Psi^{\dagger}_{k}=(a^{\dagger}_{k,A},a^{\dagger}_{k,B},a_{-k,A},a_{-k,B})$, $X(k)=2S(J-K)$, and $Y_{\pm}(k)=-2S(J\cos k\pm D\sin k)$.
We set the lattice constant $a=1$.
By using a standard Bogoliubov transformation technique \cite{colpa}, we can find $4\times4$ paraunitary matrices $\hat{Q}^{1D}_{\bm{k}}$ and $\hat{Q}^{1D\dagger}_{\bm{k}}$ that diagonalize Eq. ($\ref{1dham}$).
By performing numerical calculations, we plot magnon energies and the $z$-component magnon spin defined by Eq. ($\ref{mainresult}$) for $k$ in Figs. $\ref{fig2}$ (b) and (c).
The band structure has two splitted bands with two crossing points and a finite energy gap.
%The $z$-component magnon spin is quantized due to the absence of the U(1) symmetry breaking, while the contribution to it from each sublattice is not.
As shown in the previous section, the absence of the $U(1)$ symmetry breaking ensures that $|\bm{k},\alpha\rangle$ is an eigenstate of $S^{z}_{\mathrm{tot}}$.
Thus, the $z$-component magnon spin $S^{z}_{\bm{k},\alpha}$ is quantized, while the contribution from each sublattice does not have to be.
The fact that each contribution can be over $1$ comes from the quantum nature of the antiferromagnetic magnon \cite{quantumcorrection}.
In the upper and lower bands, the sign of $S^z_{\bm{k},\alpha}$ is changed across the band crossing points, which does not conflict with our discussion above.
This can be interpreted as a $trivial$ example of the spin-momentum locking with a collinear spin structure in momentum space.

To explore noncollinear spin structures in momentum space such as in the topological insulator surface state, 
we should consider classical ground states with noncollinear spin structures in real space [see Eq. ($\ref{mainresult}$)] \cite{noncoplanar}.

{\it Magnon spin texture in momentum space}.\textemdash 
As an example of a noncollinear structure, we consider the 120$^\circ$ structure in kagome lattice antiferromagnets, which have not only magnetic but also chemical sublattices.
We analyze the following Hamiltonian:
\begin{align}
H^{2D}=\sum_{\langle i,j \rangle}\left[\sum_aJ^aS^a_iS^a_j+\bm{D}_{ij}\cdot(\bm{S}_i\times\bm{S}_j)\right],\label{kagomeham}
\end{align}
where $J^a$ are the nearest-neighbor exchange couplings, and $\bm{D}_{ij}=\pm D\hat{z}$ is the DM vector defined in Fig. $\ref{fig1}$.
We consider the two interesting limits: \underline{(a) $J^x=J^y=J^z=J>0$, $D>0$} and \underline{(b) $J^x=J^y=J>0$, $J^z=D=0$}, both of which have classical ground states with the 120$^\circ$ structure \cite{groundstate}
and preserve the $U(1)$ symmetry around the $z$ axis.
We here choose the ground state shown in Fig. $\ref{fig1}$(b).
By mapping Eq. ($\ref{kagomeham}$) to the magnon Hamiltonian and performing the numerical Bogoliubov transformation \cite{maestro}, we plot the magnon band dispersions and $\bm{k}$-dependent magnon spin for each case in Fig. \ref{fig3} \cite{skkim}.

The band structure for the case (a) has the finite-energy flat band, which is reminiscence of the zero energy flat band in the classical spin liquid phase of the isotropic Heisenberg model.
There is one massless NG mode associated with the symmetry breaking $U(1)$$\rightarrow\{e\}$, 
and we can observe noncollinear spin structure in momentum space, as shown in Fig. $\ref{fig3}$(a). 
The norm of magnon spin is no longer quantized due to the absence of any spin rotational symmetries.
The most striking feature is that spin structures in the highest and flat bands have the winding number $Q=-2$ defined in a closed curve around a $\Gamma$ point, 
while the original spin-momentum locking in electron systems is characterized by $Q=+1$. 
Although the vector plot can depend on the choice of the ground state, all plots for $U(1)$-degenerated ground states can be identified up to overall rotation in spin space, which preserves the winding number of the vortexlike spin structures.
This model is thought to be realized in KFe$_3$(OH)$_6$(SO$_4$)$_2$ \cite{grohol,matan} except for some terms that slightly modify $\bm{S}_{\bm{k},\alpha}$ \cite{canted}.
A similar model has also been investigated in terms of a topological thermal Hall effect \cite{owerre2}.

The band structure for the case (b) also has the finite-energy flat band and one massless NG mode for the same reasons.
In addition, there are two Dirac points with a finite energy in the $K$ and $K'$ points, as shown in Fig. $\ref{fig3}$(b).
For each Dirac cone, a noncollinear spin structure characterized by $Q=+1$ is realized, which is a magnonic analogue of the topological insulator surface state.  
Note that the magnon spin-momentum locking does not require the relativistic effect, the DM interaction, while the conventional one has been found only in systems with strong relativistic effect, the spin-orbit interaction.
In magnonic systems, interesting physics can occur even in the absence of the DM interaction.
For instance, Owerre showed that the topological thermal Hall effect occurs in such a situation \cite{owerre2}.

Before ending this section, we remark on the relation between a noncollinear spin texture in the two-dimensional Brillouin zone and a mathematical theorem.
Although spin is not quantized, there is a mathematical way to characterize such spin structures, i.e., the Poincar\'e-Hopf index theorem \cite{poincare}, which states that 
\begin{align}
\chi(M)=\sum_{i}Q_i,
\end{align}
where $M$ is a compact differentiable manifold, $\chi(M)$ is the Euler characteristic of $M$, which is a typical topological invariant of the manifold \cite{poincare, euler}, $Q_i$ are the winding numbers around isolated zero points of a vector field, and the sum of indices is over all isolated zero points.
Using this theorem for the vector field $\bm{S}_{\bm{k},\alpha}$ on the two-dimensional Brillouin zone (the two-dimensional torus $\mathbb{T}^2$), we obtain the sum rule for spin-momentum locking,
\begin{align}
\sum_{i}Q_i=0,
\end{align}
where we use $\chi(\mathbb{T}^2)=0$. In the spin-momentum-locked band with $Q=-2$ discussed above, the isolated zero points are $i=\Gamma$, $K$, $K'$ with $Q_i=-2,1,1$, respectively.

$Discussion$.\textemdash 
We here briefly discuss the detection of magnon's spin angular momentum.
A current of magnon with finite spin, known as magnon spin current, can be detected by the spin pumping \cite{sandweg,kovalev1} and the spin Seebeck \cite{caloritronics,caloritronics2,kovalev2} measurements.
In conventional antiferromagnets without an external magnetic field,
net spin current in the bulk vanishes due to the degeneracy between the up and down bands \cite{ohnuma}.
The antiferromagnetic examples in Fig. $\ref{fig3}$, on the other hand, have no band degeneracy except for the crossing regions, and we can expect finite spin Seebeck signals.
However, there is no established experimental method to detect directly the momentum-dependent magnon spin, while the spin- and angle-resolved photoemission spectroscopy \cite{hasan,xlq,hoesch} enables us to detect the momentum-dependent electron spin.
To this end, we theoretically propose a setup to detect the spin-momentum locking with $Q=-2$ [Fig. \ref{fig1}(b)].
In $Q=-2$ spin structure, the spin direction of the magnon with $(k_x,0)$ is opposite to that with $(0,k_y)$.
This property would be observed as the spin flip under the magnon propagation in the specular reflection setup.
The polarized inelastic neutron scattering \cite{chatterji} is another possibility.
By investigating a change in neutron spin before and after the scattering, it is possible, in principle, to detect magnon spin-momentum locking.

In summary, we presented a theory of the magnon spin-momentum locking.
We gave conditions for magnon spin to be independent of momentum and constructed examples of spin-momentum locking by avoiding such conditions.
We find the first example of spin-momentum locking induced by spontaneous symmetry breaking.

$Acknowledgments$.\textemdash We acknowledge many fruitful discussions with Masao Ogata, Masatoshi Imada, Hiroyasu Matsuura, Yohei Yamaji, Yusuke Kousaka, Tomonari Mizoguchi,  Yohei Ema, and Yuta Kikuchi.
N. O. is supported by the Japan Society for the Promotion of Science (JSPS) through Program for Leading Graduate Schools (MERIT).
N. O. is also supported by JSPS KAKENHI (Grant No. 16J07110).


\begin{thebibliography}{9}
%magnonics
\bibitem{magnonics}V. V. Kruglyak, S. O. Demokritov, and D. Grundler, J. Phys. D $\bm{43}$, 264001 (2010).
%magnon spintronics
\bibitem{chumak}A. V. Chumak, V. I. Vasyuchka, A. A. Serga, and B. Hillebrands, Nature Phys. $\bm{11}$, 453 (2015).
%topological band theory
\bibitem{hasan} M. Z. Hasan and C. L. Kane, Rev. Mod. Phys. $\bm{82}$, 3045 (2010).
\bibitem{xlq}X.-L. Qi and S.-C. Zhang, Rev. Mod. Phys. $\bm{83}$, 1057 (2011).
\bibitem{niu} D.  Xiao,  M.-C.  Chang, and Q.  Niu, Rev. Mod. Phys. $\bm{82}$, 1959 (2010).

%magnonHall
\bibitem{fujimoto}S. Fujimoto, Phys.Rev.Lett. $\bm{103}$, 047203 (2009).
\bibitem{katsura}H. Katsura, N. Nagaosa, and P. A. Lee, Phys.Rev.Lett. $\bm{104}$, 066403 (2010).
\bibitem{onose}Y. Onose, T. Ideue, H. Katsura, Y. Shiomi, N. Nagaosa, and Y. Tokura, Science $\bm{329}$, 297 (2010).
\bibitem{owerre3}S. A. Owerre, J. Phys.:  Condens. Matter $\bm{29}$, 03LT01 (2017).
%Berry
\bibitem{shindou} R. Shindou, R. Matsumoto, S. Murakami, and J. I. Ohe, Phys. Rev. B $\bm{87}$, 174427 (2013).
%topological magnon insulator
\bibitem{lzhang}L. Zhang, J. Ren, J. -S. Wang, and B. Li, Phys. Rev. B $\bm{87}$, 144101 (2013).
\bibitem{chisnell}R. Chisnell, J. S. Helton, D. E. Freedman, D. K. Singh, R. I. Bewley, D. G. Nocera, and Y. S. Lee, Phys. Rev. Lett. $\bm{115}$, 147201 (2015).
\bibitem{owerre} S. A. Owerre, Journal  of  Applied  Physics $\bm{120}$,  043903 (2016).
%weyl magnon
\bibitem{fyli}F. Y. Li, Y. D. Li, Y. B. Kim, L. Balents, Y. Yu, and G. Chen, Nature Commun. $\bm{7}$, 12691 (2016).
%Dirac magnon
\bibitem{fransson}J. Fransson, A. M. Black-Schaffer, and A. V. Balatsky, Phys. Rev. B $\bm{94}$, 075401 (2016).

\bibitem{kittel}C. Kittel, $Introduction$ $to$ $Solid$ $State$ $Physics$ (John Wiley \& Sons, New York, 1986).

%magnon spin nernst
\bibitem{rcheng2}R. Cheng, S. Okamoto, and D. Xiao, Phys. Rev. Lett. $\bm{117}$, 217202 (2016). 
\bibitem{zyuzin} V. A. Zyuzin and A. A. Kovalev, Phys. Rev. Lett. $\bm{117}$, 217203 (2016).


%antiferro spin seebeck
\bibitem{ohnuma}
Y. Ohnuma, H. Adachi, E. Saitoh, and S. Maekawa, Phys. Rev. B $\bm{87}$, 014423 (2013).
%spin wave transistor
\bibitem{rcheng}
R. Cheng, M. W. Daniels, J.-G. Zhu, and D. Xiao, Sci. Rep. $\bm{6}$, 24223 (2016).

%winding number
\bibitem{winding}
Although it is not a mathematically rigorous definition, we would like to give the intuitive meaning of the winding number below. Let $\bm{v}(\bm{x})$ be a vector field with an isolated zero $\bm{x}_0$. The winding number is the total number of counterclockwise turns experienced by $\bm{v}(\bm{x})$ after completing the counterclockwise motion along the closed curve $\gamma$ around $\bm{x}_0$.
Positive winding number means the counterclockwise turns of $\bm{v}(\bm{x})$ and conversely, negative winding number means the clockwise turns.
The winding number can be calculated by the formula:
$Q=1/(2\pi)\oint_{\gamma} ds (\bm{n}\times\partial_s\bm{n})_z$, where $\bm{n}=\bm{v}_{\parallel}/|\bm{v}_{\parallel}|$, and $\bm{v}_{\parallel}$ is the in-plane projection of $\bm{v}$.



\bibitem{colpa}
 J. H. P. Colpa, Physica A $\bm{93}$, 327 (1978).
\bibitem{magmageffect}
In this paper, we treat the magnets in the semi-classical picture ($S_0\gg1$).
It would be an interesting future work to include interaction effects in $S_0\sim1$ systems such as nonperturbative damping discussed in 
A. L. Chernyshev and P. A. Maksimov, Phys. Rev. Lett. $\bm{117}$,187203 (2016).
\bibitem{rotsym}

In SU(2) symmetric systems, we can not define the spin-momentum locking due to the spin degeneracy.
In U(1) symmetric systems, all Bloch states in the Brillouin zone have the common quantized axis, and spin structures can not be interesting.
\bibitem{supplemental}
See the Supplemental Material for the details of calculation.
\bibitem{triangular}
Note that spins on different magnetic sublattices in spin Hamiltonians without chemical sublattices such as a triangular lattice antiferromagnet contribute to $\bm{S}_{\bm{k},\alpha}$ with the equal weight, and cancel out each other.

\bibitem{watanabe} 
In this simple symmetry breaking, the number of massless NG bosons $n_{\mathrm{NGB}}$ is given by dim U(1)/$\{e\}$=1.
The formula to calculate $n_{\mathrm{NGB}}$ is given in H. Watanabe and H. Murayama, Phys. Rev. Lett. $\bm{108}$, 251602 (2012).




\bibitem{quantumcorrection}
In usual context, this is known as the quantum correction to the classical antiferromagnetic ground state.
See P. Fazekas, $Lecture$ $Notes$ $on$ $Electron$ $Correlation$ $and$ $Magnetism$
(World Scientific, Singapore,1999).
\bibitem{noncoplanar}
Although we focus on the non-collinear spin structure, there is no reason to forbid the non-coplanar spin structure in momentum space, which could be realized in non-coplanar magnets.

\bibitem{groundstate}
Strictly speaking, there is another ground state, $\sqrt{3}\times\sqrt{3}$ structure, for the case (b).
In the presence of small but finite DM interaction, which exists in realistic materials, we do not have to consider this degeneracy. 
\bibitem{maestro}
A. D. Maestro and M. Gingras, J. Phys. Cond. Matt. $\bm{16}$, 3399 (2004).

\bibitem{skkim}
Note that the concept of magnon spin is different from the h-vector, which is closely related to the topological property, plotted such as in
S. K. Kim, H. Ochoa, R. Zarzuela, and Y. Tserkovnyak, Phys. Rev. Lett. $\bm{117}$, 227201 (2016).

\bibitem{grohol}
D. Grohol, K. Matan, J.-H Cho, S.-H. Lee, J. W. Lynn, D. G. Nocera and Y. S. Lee, Nat. Mater. $\bm{4}$, 323 (2005).
\bibitem{matan}
K. Matan, D. Grohol, D. G. Nocera, T. Yildirim, A. B. Harris, S. H. Lee, S. E. Nagler, and Y. S. Lee, Phys. Rev. Lett. $\bm{96}$, 247201 (2006).
\bibitem{canted}
As mentioned in Refs. \cite{grohol} and \cite{matan}, the ground state spins are canted slightly out of the kagome plane.
Thus, the magnons have small z-component spin angular momentum accordingly.

\bibitem{owerre2}
S. A. Owerre, Phys. Rev. B $\bm{95}$, 014422 (2017).
%\bibitem{tomiyoshi}
%S. Tomiyoshi and Y. Yamaguchi, J. Phys. Soc. Jpn. $\bm{51}$, 2478 (1982).
%\bibitem{nagamiya}
% T.  Nagamiya,  S.  Tomiyoshi,   and  Y.  Yamaguchi,  Solid  State Commun. $\bm{42}$, 385 (1982).
%\bibitem{hchen}
%H. Chen, Q. Niu, and A. H. MacDonald, Phys. Rev. Lett. $\bm{112}$, 017205 (2014).

\bibitem{poincare}
 V. Guillemin and A. Pollack, $Differential$ $topology$ (American Mathematical Society, 2010).
\bibitem{euler}
M. Nakahara, $Geometry,$ $Topology,$ $and$ $Physics$ (A. Hilger, 1990).
%spin pumping
\bibitem{sandweg}C. W. Sandweg, Y. Kajiwara, A. V. Chumak, A. A. Serga, V. I. Vasyuchka, M. B. Jungfleisch, E. Saitoh, and B. Hillebrands, Phys. Rev. Lett. $\bm{106}$, 216601 (2011).
\bibitem{kovalev1}A. A. Kovalev, V. A. Zyuzin, and B. Li, Phys. Rev. B $\bm{95}$,165106 (2017).
\bibitem{kovalev2}A. A. Kovalev and V. A. Zyuzin, Phys. Rev. B $\bm{93}$, 161106 (R) (2016).
%spin caloritronics
\bibitem{caloritronics}G. E. W. Bauer, A. H. MacDonald, and S. Maekawa (Eds), $Spin$ $Caloritronics$, $Special$ $Issue$ $of$ $Solid$ $State$ $Communications$ (Elsevier, 2010). 
\bibitem{caloritronics2}G. E. W. Bauer, E. Saitoh,   and B. J. van Wees, Nature materials $\bm{11}$, 391 (2012).


\bibitem{hoesch}
M. Hoesch, M .Muntwiler, V. N. Petrov, M. Hengsberger, L. Patthey, M. Shi, M. Falub, T. Greber, and J. Osterwalder, Phys. Rev. B $\bm{69}$, 241401(R) (2004).
\bibitem{chatterji}
T. Chatterji, $Neutron$ $Scattering$ $from$ $Magnetic$ $Materials$ (Elsevier, Amsterdam, 2006).
\end{thebibliography}
\end{document}

\preprint{APS/123-QED}
\title{Supplemental Material for\\ ``Magnon Spin-Momentum Locking: Possible Realizations in Non-collinear Antiferromagnets''}

\author{Nobuyuki Okuma}
\email{okuma@hosi.phys.s.u-tokyo.ac.jp}
 %\altaffiliation{Department of Physics, University of Tokyo, Hongo 7-3-1, 113-0033, Japan}%Lines break automatically or can be forced with \\
\affiliation{%
 Department of Physics, University of Tokyo, Hongo 7-3-1, Tokyo 113-0033, Japan
}%

\maketitle
\section{Physical interpretation of magnon spin}
\begin{figure}[]
\begin{center}
\includegraphics[bb=0 0 578 289,width=7cm,angle=0,clip]{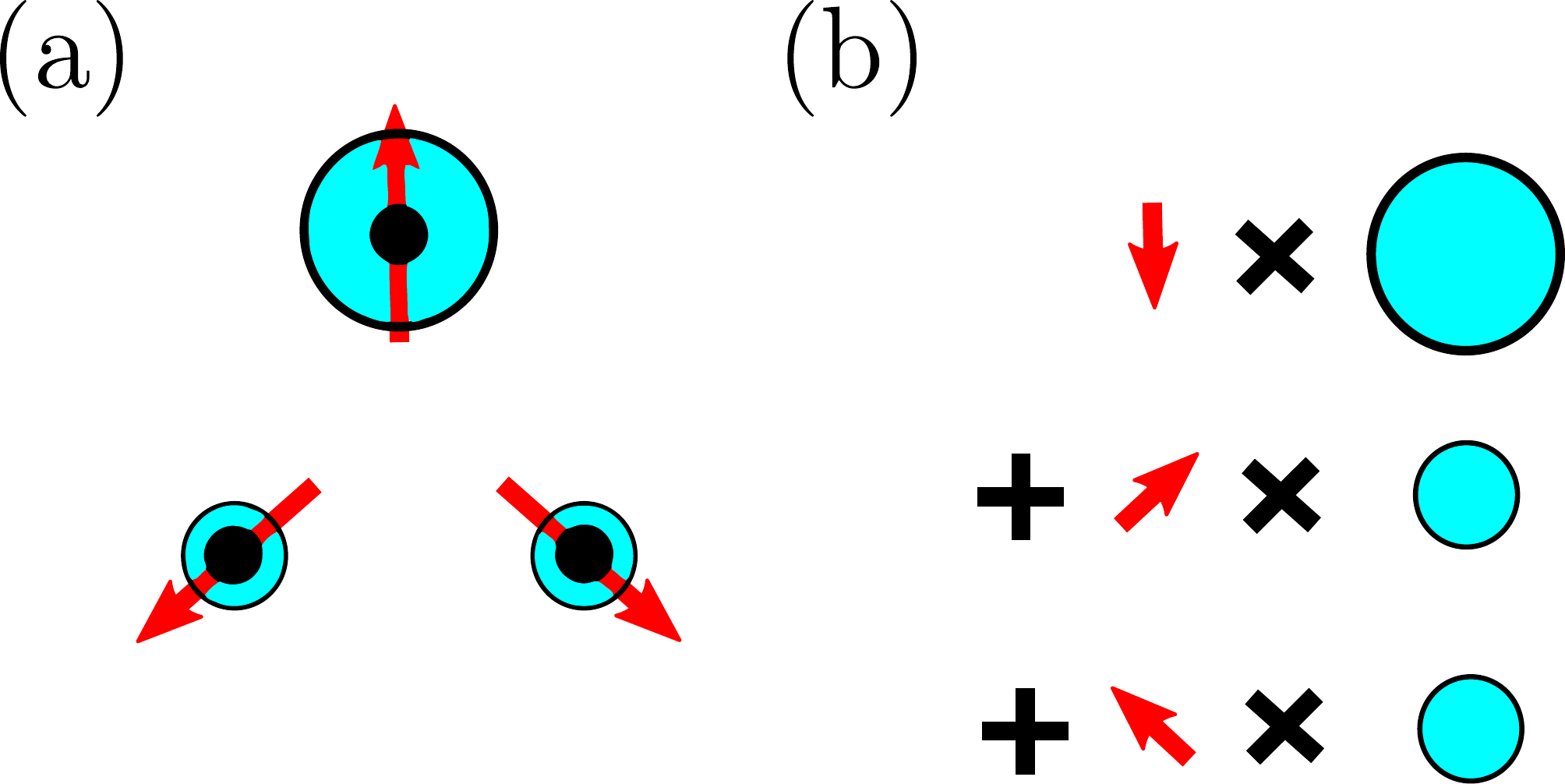}
\caption{Schematic pictures of (a) magnon probability distribution function and (b) magnon spin.}
\label{fig4}
\end{center}
\end{figure}

The physical interpretation of the formula (7) is as follows.
The Holstein-Primakoff creation operator $a^{\dagger}_{\bm{R},i}$ is physically equivalent to the spin lowering operator with respect to the classical spin direction $\bm{M}^z_i$.
In the presence of one Holstein-Primakoff boson at a sublattice $i$, the change of the total spin is $-\bm{M}^z_i$.
Since a one-magnon state $|\bm{k},\alpha\rangle$ is described as the superposition of one Holstein-Primakoff boson states [see Fig. $\ref{fig4}$],
the $\bm{k}$-dependent magnon spin can be interpreted as
\begin{align}
\sum_i&[-(\mathrm{classical\ spin\ unit\ vector\ at\ }i )\notag\\
&\times(\mathrm{probability\ distribution\ function\ at\ }i)].
\end{align}
Note that the total probability does not have to be 1 due to the paraunitarity of $\hat{Q}$.

\section{Total spin operators for rotational symmetric hamiltonians}

We here construct the total spin operators $S^{a}_{tot}$ that satisfy $[H,S^{a}_{tot}]$ in terms of magnon field operators $(b,b^{\dagger})$ upto the second order.

Since constant terms trivially commute with the Hamiltonian, we first consider the first-order terms of $(b,b^{\dagger})$.
The commutator $[H,b_{\bm{k},\alpha}^{(\dagger)}]$ is calculated as
\begin{align}
[H,b_{\bm{k},\alpha}^{(\dagger)}]=\left[\sum_{\bm{k'},\beta}\epsilon_{\bm{k},\beta}b^{\dagger}_{\bm{k'},\beta}b_{\bm{k'},\beta},b_{\bm{k},\alpha}^{(\dagger)}\right]
=\mp\epsilon_{\bm{k},\alpha} b^{(\dagger)}_{\bm{k},\alpha}.
\end{align}
To satisfy $[H,S^{a}_{tot}]$, the total spin operator can only include the zero-energy magnon modes, or equivalently, the massless NG modes.
Thus, the first order part of total spin operator $S^{a,1\mathrm{st}}_{tot}$ is given by
\begin{align}
S^{a,1\mathrm{st}}_{tot}=\sum_m \left(C^a_mb_{NG,m} +(C^a_m)^*b^{\dagger}_{NG,m}   \right),
\end{align}
where $C^a_m$ are complex numbers, and ($b_{NG,m},b^{\dagger}_{NG,m}$) are field operators of massless NG modes.
We here use the fact that $S^{a,1\mathrm{st}}_{tot}$ should be Hermitian operators.

Next, we consider the second-order terms.
For one-magnon states $|\bm{k},\alpha\rangle$ and $|\bm{k},\beta\rangle$, the following relation holds:
\begin{align}
0=\langle\bm{k},\alpha| [H,S^{a}_{tot}]|\bm{k},\beta\rangle=(E_{\bm{k},\alpha}-E_{\bm{k},\beta})\langle\bm{k},\alpha|S^{a}_{tot}|\bm{k},\beta\rangle.
\end{align}
For the systems without degeneracy ($E_{\bm{k},\alpha}\neq E_{\bm{k},\beta}$), we obtain
\begin{align}
\langle\bm{k},\alpha|S^{a}_{tot}|\bm{k},\beta\rangle=\delta_{\alpha\beta}S^{a}_{\bm{k},\alpha},
\end{align}
and the second order part of total spin operator $S^{a,2\mathrm{nd}}_{tot}$ is given by
\begin{align}
S^{a,2\mathrm{nd}}_{tot}=\sum_{\bm{k},\alpha}S^a_{\bm{k},\alpha}b^{\dagger}_{\bm{k},\alpha}b_{\bm{k},\alpha}.
\end{align}
For the systems with degeneracy, we can always choose the basis to diagonalize $\langle\bm{k},\alpha|S^{a}_{tot}|\bm{k},\beta\rangle$ by a proper unitary transformation.
This statement does not mean that $\langle\bm{k},\alpha|S^{a}_{tot}|\bm{k},\beta\rangle$ can be simultaneously diagonalizable for every $a=x,y,z$.
However, it is not important for our discussion in the main text because we do not use the bosonic representations of $S^x_{tot},S^y_{tot},S^z_{tot}$ simultaneously.

In summary, we obtain the following bosonic representations of the total spin operators upto the second order:
\begin{align}
S^a_{tot}=&(Const.)+\sum_{\bm{k},\alpha}S^a_{\bm{k},\alpha}b^{\dagger}_{\bm{k},\alpha}b_{\bm{k},\alpha}\notag\\
&+\sum_m \left(C^a_mb_{NG,m} +(C^a_m)^*b^{\dagger}_{NG,m}   \right).
\end{align}